\documentclass[aps,pra,twocolumn,10pt]{revtex4-1}

\usepackage{graphicx}
\usepackage{amsmath}
\usepackage{mathptmx}   
\usepackage{amssymb}
\usepackage{pslatex}	    

\begin{document}

\title{Rapidly reconfigurable optically induced photonic crystals in hot rubidium vapor}

\author{Bethany Little$^1$}
	\email{bethany@pas.rochester.edu}
\author{David J. Starling$^{1,2}$}	
\author{John C. Howell$^1$}
\author{Raphael Cohen$^3$}
	\email{raphael.cohen3@mail.huji.ac.il}
\author{David Shwa$^3$}
\author{Nadav Katz$^3$}

\affiliation{$^1$Department of Physics and Astronomy, University of Rochester, Rochester NY 14620}
\affiliation{$^2$Division of Science, Pennsylvania State University, Hazleton, PA 18202}
\affiliation{$^3$The Racah Institute of Physics, The Hebrew University, Jerusalem 91904, Israel}

\date{\today}

\begin{abstract}
Through periodic index modulation, we create two different types of photonic structures in a heated rubidium vapor for controlled reflection, transmission and diffraction of light. The modulation is achieved through the use of the AC Stark effect resulting from a standing-wave control field.  The periodic intensity structures create translationally invariant index profiles analogous to photonic crystals in spectral regions of steep dispersion.  Experimental results are consistent with modeling.
\end{abstract}


\maketitle

\section{Introduction}


All-optical structures which can be dynamically controlled are an important resource for many applications, such as ultra-fast optical switches \cite{Asakawa} or nonlinear processes in waveguides \cite{Vudyasetu}. The photonic crystal \cite{Sajeev1987, Joannopoulos} is one such all-optical structure, in which light propagating in a periodic dielectric medium has a band structure of allowed and forbidden frequencies. Another useful structure is the thick Bragg grating or more generally the volume hologram which may be used for information storage \cite{Kogelnik}. By carefully constructing a dielectric medium, light can be guided and shaped for use in information technology and other applications \cite{Photonics}.

Much progress has been made in this field \cite{Joannopoulos}, including the dynamical control of the structure by using the electro-optic effect \cite{Scrymgeour} and fast pulses \cite{Almeida:2004fk, Huang, Wen2012}.  However, as Artoni and La Rocca point out \cite{Artoni2006}, all of these schemes are limited by the original specifications of the structure; while the index of the medium may be rapidly changed, the spatial structure cannot be modified quickly.

One interesting solution to this problem is to ``write'' a structure onto a medium using interfering laser beams, such as in Ref.~\cite{Ling}.  By making use of electromagnetically induced transparency (EIT), a periodic structure with alternatively transparent and opaque regions can be created with counter-propagating beams. It has been shown that this effect could also be used to create structures that behave like photonic crystals \cite{Andre} which can be modified in time \cite{Gao2010}.  Notably, in addition to being rapidly reconfigurable, these structures also have high spatial fidelity, avoiding the local disorder problems of solid crystals \cite{Artoni2006}.

However, all of these systems which utilize EIT are limited by coherence times and are more readily realizable in cold atoms than at room temperature owing to the ballistic or diffusive behavior of the atoms. Another possibility is to create structures using optical pumping \cite{Korneev}, but the bandwidth will be limited by the decay time.

In this paper we propose and demonstrate photonic structure behavior within a cell of hot rubidium gas by using the AC Stark effect.  In addition to avoiding the use of cold atoms, theses structures occur over large control beam detunings (potentially answering some questions posed in Ref.~\cite{Brown}) and can be modified at a rate exceeding Ref.~\onlinecite{Gao2010}. This leads to many interesting possibilities for future work, such as rapidly creating a cavity at the location of a short pulse of light to trap it, quickly reflecting a beam of a certain frequency, or making use of slow light such as in \cite{Baba2008}.

In the following section we introduce the theory based on the AC Stark effect in rubidium gas, and in sections \ref{sec:Photonic} and \ref{sec:Bragg} we describe two different experiments, both of which demonstrate Stark effect induced structural behavior.  Finally, conclusions and future work are discussed in section \ref{sec:Conclusions}.

\section{Theory}\label{sec:theory}

\begin{figure*}
	\begin{center}
		\includegraphics[scale=1]{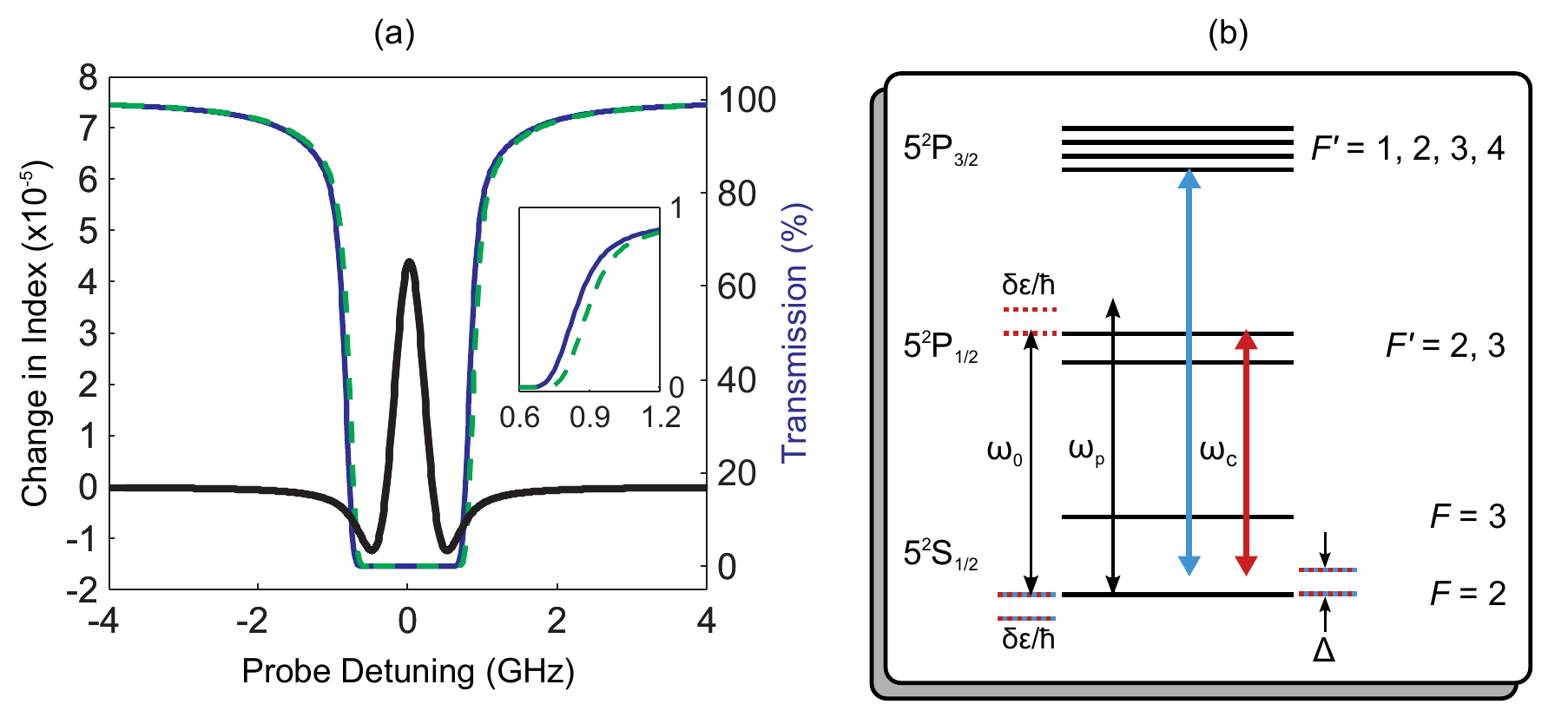}
		\caption{\label{fig:the} (Color online) (a) The modification of the refractive index and transmission for a typical experimental setup (control beam: $P = 30$ mW, a beam waist of $810 \times  85$ $\mu$m$^2$, $T =85^\circ$C and a cell length of 7.5 cm). The transmission curves for a probe beam with (dashed green) and without (solid blue) the control field present are shown along with the resulting change in the real part of the refractive index (solid black). (b) The relevant levels for rubidium 85 are shown. The common ground state ($5^2 S_{1/2}$ $F = 2$) is Stark-shifted by the control beam either 780 nm (blue, solid) or 795 nm (red, dashed). The magnitude of the Stark shift is included as $\delta \epsilon/\hbar$ for two levels.}
	\end{center}
\end{figure*}

The phase shift and absorption caused by an atomic transition are related to the real and imaginary parts of the susceptibility of that transition \cite{Siegman}. A sufficiently strong control laser will, via the AC Stark effect, cause a shift in the transition frequency. The resulting change in the susceptibility will depend upon the intensity of the control laser and the temperature and density of the atoms. This change in the susceptibility causes an adjustment to the refractive index and transmission; the expected modification for a typical setup is shown in Figure \ref{fig:the}a. For a given propagation length, it can be seen that there is a trade-off between absorption and change in refractive index. For a fixed control beam intensity (and hence a fixed AC Stark shift), a greater refractive index change can be achieved by increasing the density of the rubidium atoms; however, this in turn leads to greater absorption. 

The AC Stark shift caused by a control field of frequency $\omega_c$ may be calculated by considering the field to be constant over the region of a given atom and using the standard Hamiltonian for a two-level atom with resonance frequency $\omega_0$ interacting with an electric field given by $\vec{E}=\frac{1}{2}(\vec{E_c}e^{-i\omega_ct}+\vec{E_c^*}e^{i\omega_c t})$.  We assume the laser is near resonance and use the rotating wave approximation, where the Hamiltonian becomes
\begin{equation}
	\mathcal{H_R}=-\hbar
		\begin{pmatrix}
			0 & \Omega/2 \\
			\Omega^*/2 & \Delta
		\end{pmatrix},
\end{equation}
with Rabi frequency $\Omega=\vec{d}\cdot \vec{E_c}/\hbar$, detuning $\Delta=\omega_c-\omega_0$ and $\vec{d}$ is the transition dipole moment. The shifted levels can be found from the eigenvalues of $\mathcal{H_R}$.

If the control field is chosen to interact only with the ground state (which is common to both the probe and control fields), as in the 780 nm field of Figure \ref{fig:the}b, then the expected shift of the probe resonance can be found by considering the shift on just the lower level. This is given by 

\begin{equation}\label{eq:shift}
	\delta \epsilon= -\frac{\hbar}{2}\left(\Delta-\sqrt{\Delta^2+|\Omega|^2}\right).
\end{equation}
For large detuning, this becomes
\begin{equation}\label{eq:largedetuning}
	\delta \epsilon \approx -\frac{\hbar|\Omega|^2}{4\Delta}.
\end{equation}

By interfering the control beam with itself, a periodic standing wave can be created within a heated cell of rubidium. This standing wave provides a periodic intensity and thus a periodic change in refractive index, analogous to a photonic structure, but created in a hot cloud of rubidium gas.

For small AC Stark shifts the change in refractive index $\Delta n$ will be proportional to the shift in susceptibility; but the susceptibility is proportional to the AC stark shift $\delta \epsilon/\hbar$, hence linear in control intensity and inversely proportional to the control detuning [via Eq.~\eqref{eq:largedetuning}]. The efficiency of such structures will be proportional to the square of this refractive index contrast [Ref.~\onlinecite{Kogelnik}, Eq.~(45)]; that is to say, the efficiency has the behavior
\begin{equation} \label{eq:efficiency}
	e \propto I_c^2 \,\,\text{ and }\,\, e \propto \frac{1}{\Delta^2},
\end{equation}
where $I_c$ is the intensity of the coupling field. It is therefore advantageous to increase the coupling field intensity and reduce the detuning (while still avoiding other processes such as optical pumping). 

\section{Photonic Band Gap}
\label{sec:Photonic}

\subsection{Experiment}
\begin{figure}
	\begin{center}
		\includegraphics[scale=1]{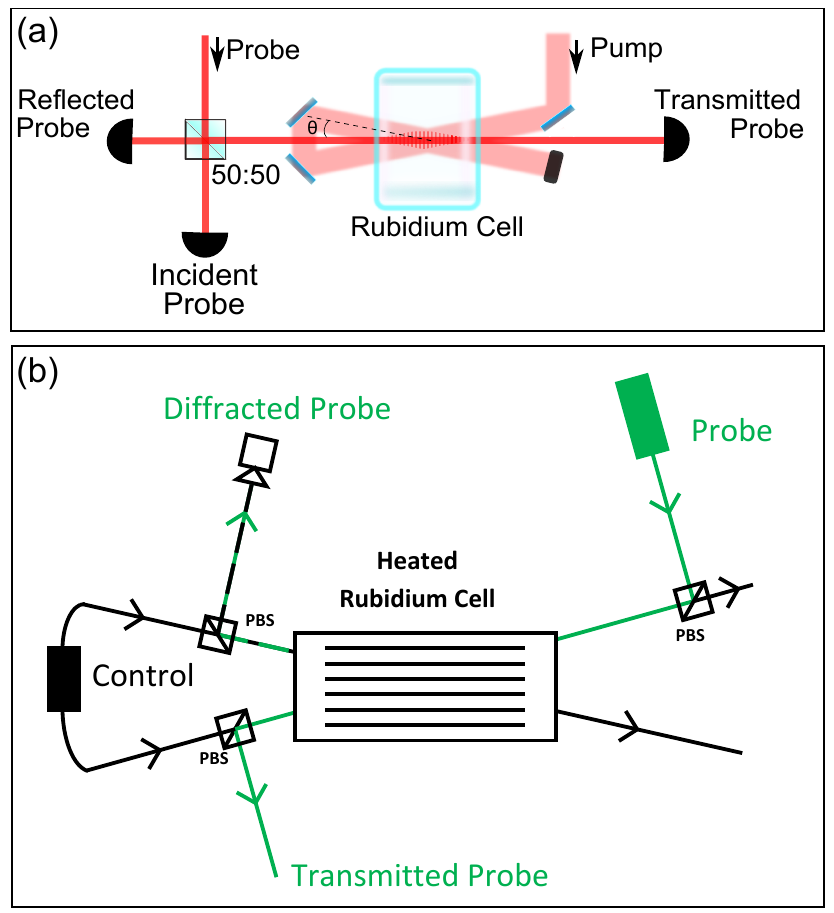}
		\caption{\label{fig:exp} (Color online) (a) Schematic of the photonic band gap experimental setup of section III. Probe light at 795 nm is incident on a rubidium vapor cell and appropriate detectors.  A Stark-shifting control beam at 780 nm and detuning $\Delta$ of at least 1 GHz is set up in a crossed configuration in order to create an interference pattern with a periodicity that creates a band gap in the medium for the probe beam. (b) Schematic of the Bragg grating experimental setup of section IV. Polarizing beamsplitters (PBS) are used to combine and separate the control and probe fields.}
	\end{center}
\end{figure}	

One such structure can be induced by using a crossed beam setup, shown in Figure \ref{fig:exp}a, which creates a set of disallowed frequencies for a probe beam.  We chose to use a crossed-beam setup rather than a collinear design to enhance filtering, to avoid coherent effects, and to prevent the control beam from creating a band gap for itself.  We choose the angles between the beams such that the distance $\Delta p$ between nodes of the control beam is half the wavelength of the probe beam.  Since the index contrast $\Delta n = n_1-n_2$ between the node and antinode of the control field is small, this creates the equivalent of a quarter-wave stack, with $n_1a_1\approx n_2 a_2$ and $a_1=a_2=\Delta p/2$.

For two control beams with wave numbers $k_1=k_2=2\pi/\lambda_c$ we can calculate the required angle $2\theta$ between the  beams for a band gap to occur.  Treating the two crossing control beams as plane waves of the form $\vec{E}_j=\vec{A_j}e^{i(\vec{k}\cdot \vec{r}-\omega t)}+c.c.$ ($j=\{1,2\}$) and assuming the amplitudes of each beam $A_1=A_2=A$, the intensity as a function of distance $x$ away from their crossing point for a given angle $\theta$ is
\begin{equation}
	I=2A^2\{1-\cos [2\theta] \cos [2kx \cos(\theta)] \}.
\end{equation}
Solving for the spacing $\Delta p$ between peaks, we find that $\Delta p = \lambda_c /[2 \cos (\theta)]$, where $\lambda_c$ is the wavelength of the control field. Letting the spacing be half the wavelength of the probe field ($\lambda_p$) such that $\Delta p = \lambda_p/2$, we find the simple condition
\begin{equation}\label{eq:angle}
	\cos (\theta) = \lambda_c/\lambda_p.
\end{equation}
For $\lambda_c= 780$ nm and $\lambda_p = 795$ nm, this gives $\theta=11.2^\circ$.

The experimental setup consists of a 780 nm external cavity diode laser that produces a horizontally polarized standing wave in a 1 mm long cell with naturally abundant rubidium. A vertically polarized probe beam generated by a 795 nm external cavity diode laser is incident on a 50:50 beamsplitter.  Half of the light passes through the cell to a detector, where the absorption spectra is measured.  The other half is measured at a second detector so that percentage-of-input calculations can be made.  Reflected probe light passes back through the beamsplitter to a third detector.

Absorption, transmission and reflection spectra of the probe beam were recorded for various control beam frequencies.  The percent reflected was calculated by subtracting background from the reflected power and dividing by the input power.  The control beam was detuned from resonance by at least 1 GHz to limit optical pumping.  As expected, the frequency of the band gap depended on the frequency of the control light, but reflection was still present for detunings exceeding 50 GHz. Temperature adjustments were used to optimize the optical depth. The angle between the control beams was tuned to maximize reflection for a given probe frequency, consistent with Eq.~\eqref{eq:angle}. 

\subsection{Results}
\label{sec:Photonic_results}

The results of the experiment are consistent with the model of a periodic index contrast induced by the AC Stark effect.  This is evident through observations of certain disallowed probe frequencies (exhibited by reflection from the gas). The disallowed probe frequencies depend on both the angle and the frequency of the control field, despite the lack of coherence present in the experiment.

Eq.~\eqref{eq:shift} can be used to find the expected Stark shift from the control field. For a control power of 120 mW, a gaussian beam diameter of 600 $\mu$m and a detuning of 2 GHz, the predicted shift is $5.7\times10^{7}$ Hz. Based upon probe scans with and without the control field, the average shift was found to be $2\times 10^{7}$ Hz. However, this is only an approximation that does not take into account the distortion of the transmitted field plot when the standing wave control field is present.

\begin{figure}
	\begin{center}
		\includegraphics[scale=1]{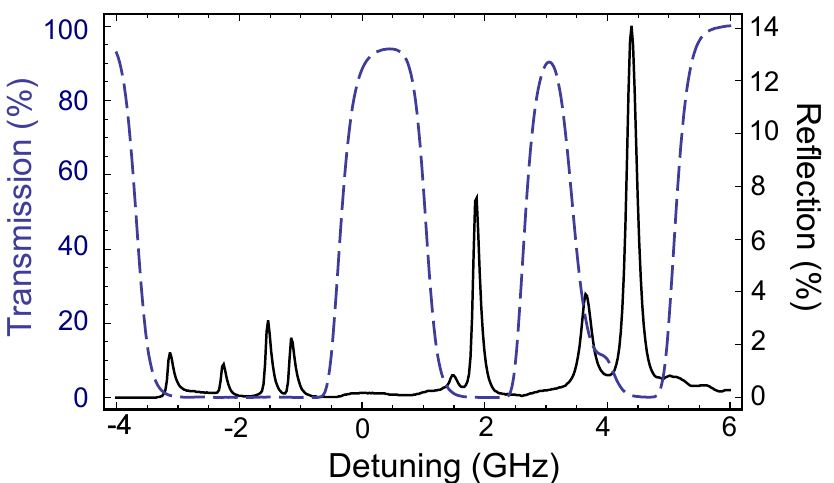}
		\caption{\label{fig:data1} (Color online) Plot of transmitted (dashed) and reflected (solid) probe light.  Since the index contrast created by the Stark effect is greatest near resonance, the band gap is more dominant in these regions.}
	\end{center}
\end{figure}

The reflection and transmission spectra of the probe are shown in  Figure \ref{fig:data1}.  The steepest dispersion occurs in regions near resonance; therefore, the reflectivity is strongest there, consistent with Figure \ref{fig:the}. This is also where the greatest absorption occurs, accounting for the low percentage of reflected light. Depending on the frequency and angle of the control beams, the reflection occurs strongly at the corresponding probe frequencies. In the data shown, the angle was tuned so that the reflected peak was maximized for the $D_1$ transition of $^{87}$Rb, from the $F=1$ ground state. It is not surprising that the effect occurs at the other resonances (although more weakly), since according to Eq.~\eqref{eq:shift} a Stark shift is still present for nonzero detuning.  It is also important to note that the removal of one of the control beams results in no frequency-dependent reflections; this behavior supports our conclusion that the standing wave is responsible for the band gap.


\begin{figure}
	\begin{center}
		\includegraphics[scale=1]{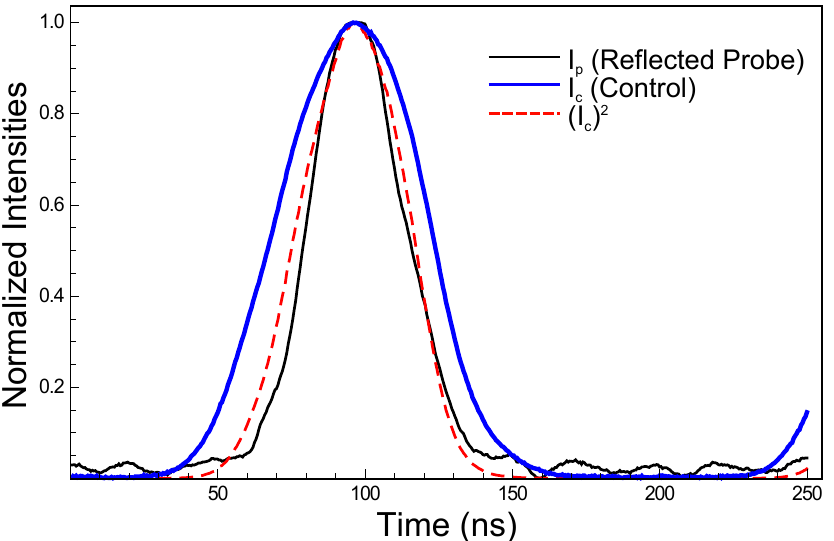}
		\caption{\label{fig:datatime} (Color online) Plot of probe (black, thin) and control field (blue, thick) normalized intensities along with the control field squared (red, dashed) during pulsed operation. A rapid response of the band gap effect to the turn-on of the control field is evident.}
	\end{center}
\end{figure}	

The time response of the system was then investigated by pulsing the control field and observing the reflected light.  Figure \ref{fig:datatime} shows a plot of the amplitude of both the control field and the reflected probe field as a function of time.  The peak of the reflected pulse occurs at the same time as the peak of the control field pulse, within the precision of our measurement ($\approx 5$ ns). Faster pulsing and detecting apparatus would be required to detect a delay in the response of the band gap, which should be proportional to the Rabi frequency of the control field. This will be the subject of future work. However, we note here that the delay of the reflected pulse in Ref.~\onlinecite{Gao2010} was on the order of a microsecond.

It may also be noted that the width of the reflected probe pulse is shorter than the control field pulse (thin black versus thick blue traces in Figure \ref{fig:datatime}). However, by comparing the width of the square of the intensity to the width of the reflected probe pulse (dashed red versus thin black traces in Figure \ref{fig:datatime}), we find the results consistent with the discussion of Eq.~\eqref{eq:efficiency}.

This first demonstration is the all-optical equivalent of a quarter wave stack in a dielectric medium. In the following section, we turn to a second arrangement where we again employ the AC Stark effect, but in a co-propagating configuration leading to a thick Bragg grating. 

\section{Bragg Grating}
\label{sec:Bragg}

\subsection{Experiment}

To create the Bragg grating we interfere two nearly co-propagating beams taken from the same laser source, as shown in Figure \ref{fig:exp}b. A standing wave is produced in the plane perpendicular to the average direction of propagation of the two control beams. If the control and probe wavelengths are close, then the Bragg angle for such a grating is identical to the angle of incidence of the control beam. The path of the probe and its diffraction therefore lie along the paths of the control beams. Thus, we are able to use orthogonal polarizations to separate the probe from the control. 
In contrast to the experiment of section \ref{sec:Photonic}, here we use an isotopically pure $^{85}$Rb vapor cell of length 7.5 cm. Also, in this configuration, it is desirable to use nearly degenerate control and probe fields (in this case the $D_1$ line of rubidium).

We simulate the effect of this grating by first calculating the expected change in the refractive index caused by the control field using classical susceptibility equations, calculated from \cite{Siegman}, with the added AC Stark shift of Eq.~\eqref{eq:largedetuning} and taking into consideration hole burning and optical pumping caused by the probe. These results, along with experimental measurements of the beam sizes and angles, are then incorporated into a simulation of the 3D propagation of the electromagnetic field. 

If the intensity profile of the interfering control beams $I(x,y,z)$ is known, the accumulated phase difference of the probe field can be expressed as
\begin{equation}\label{eq:phaseshift}
\phi (x,y,z) = \frac{2\pi}{\lambda}\delta n I(x,y,z) z+i[\alpha +\delta \alpha I(x,y,z)]z,
\end{equation}
where $n$ is the real part of the total index of refraction $n + i n''$, $\alpha$ is the absorption coefficient (given by $\alpha = 2 \pi \nu n''/c$), and $\delta n$ and $\delta \alpha$ are the corresponding changes. The intensity of the diffracted and transmitted probe field was found by summing the absolute square of the field over the pixels in the appropriate diffraction order.

\begin{figure}
	\begin{center}
		\includegraphics[scale=1]{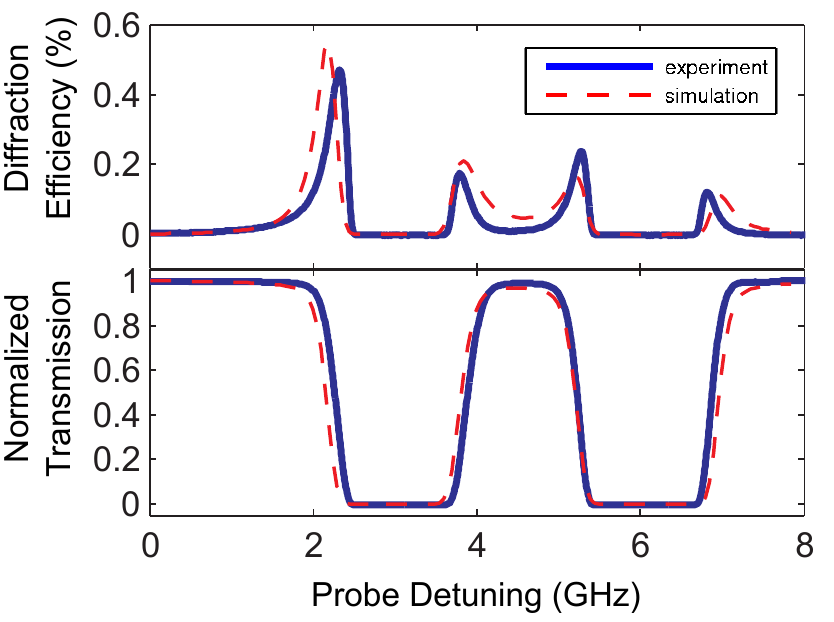}
		\caption{\label{fig:probe} (Color online) (a) Plot of measured (blue) and simulated (red, dashed) diffracted probe light.  (b) Measured (blue) and simulated (red, dashed) transmitted probe. Probe detuning is from the $D_1$ line of the $^{87}$Rb  $F = 1 \rightarrow F' = 2$ transition. Cell and control parameters are same as in Figure \ref{fig:the} and the grating spacing is 55 $\mu$m.}
	\end{center}
\end{figure}

\subsection{Results}
\label{sec:Bragg_results}

We first measure the diffracted signal, shown in Figure \ref{fig:probe}, by tuning the control beam to the $D_1$ line of the $^{87}$Rb $F=1\rightarrow F'=2$ transition, 2.7 GHz from the closest  $^{85}$Rb transition; this ensures that there is no optical pumping. (Again, the reflected probe field is not present when we block either of the control beams, thus verifying that our signal is due to the induced grating). We see that the diffracted signal is strongest on the edges of the absorption as expected (see Figure \ref{fig:the}), and is observed to be largest on the $^{85}$Rb $D_1$ $F = 2$ transition where there is the smallest detuning. The results of the simulation confirm that the grating is produced by the AC Stark shift.

\begin{figure}
	\begin{center}
		\includegraphics[scale=1]{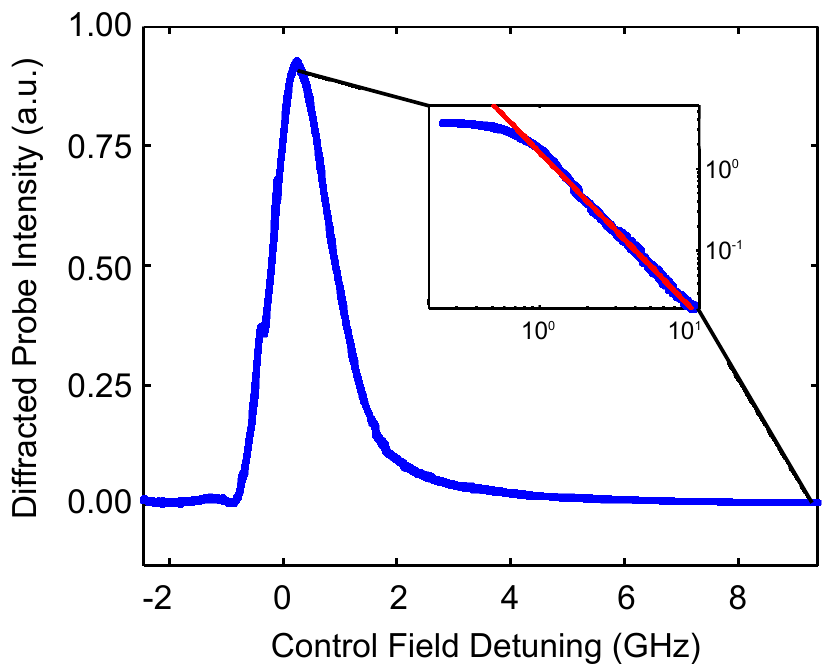}
		\caption{\label{fig:pump} (Color online) Plot of measured diffracted probe with change in the detuning of the control field (blue). The inset shows a log-log plot of the measured data (blue) and a straight line fit (red) with a slope of $-1.99 \pm 0.02$. The control field detuning is measured from the $D_1$ line of the $^{85}$Rb $F = 2 \rightarrow F' = 3$ transition. }
	\end{center}
\end{figure}

Finally, we vary the control field frequency while holding the probe at a constant detuning of 2.2 GHz from the $D_1$ line of the $^{87}$Rb $F = 1 \rightarrow F' = 2$ transition. The diffracted probe signal is shown in Figure \ref{fig:pump} and is visible for pump detunings up to 10 GHz. The diffracted probe signal falls off with an inverse square law relationship consistent with an AC Stark effect induced grating as predicted by the theory [Eq.~\eqref{eq:efficiency}]. This behavior is shown in the inset of Figure \ref{fig:pump}, where the slope is $-1.99 \pm 0.02$. When the control is close to the $^{85}$Rb $D_1$ $F=2$ transition, optical pumping dominates as expected. On the other side of the transition the new dressed state caused by the control beam causes the probe to lie in a place of very high absorption; thus no diffracted signal is observed at this probe frequency.


\section{Conclusions }
\label{sec:Conclusions}

Both systems show optically induced structures that are only effective when the interfering control fields are present. It is clear from these results that the AC Stark effect is the cause of these structures and that the response time is $\lesssim 5$ ns.

The results of Section \ref{sec:Photonic_results} are consistent with the model of a periodic index contrast induced by the AC Stark effect, which creates the photonic band gaps seen in Figure \ref{fig:data1}. The results of Section \ref{sec:Bragg_results} demonstrate that we are able to very closely simulate these processes, and that to fully understand the behavior of these structures one needs to take into account the change in both the real and imaginary parts of the susceptibility caused by the AC Stark effect. In future work, we plan to extend the periodic structure into two dimensions. This could be achieved with the use of a digital micro-mirror device capable of generating periodic patterns resembling a 2D photonic crystal. This pattern could be easily manipulated at high speed, effectively creating a fast, controllable optical circuit.

In summary, we have shown that a variety of structures may be set up in hot rubidium gas by use of the AC Stark effect. By turning the control fields on (off) we can enable (disable) the photonic band gap, thereby producing a fast optical switch. Importantly, the control and probe fields can be derived from different sources and the atomic cloud is heated; this is a vast simplification over designs that utilize EIT and cold atoms. These results open up many interesting possibilities in quantum computing and alternatives to photonic crystal technology. 


\section*{Acknowledgements}

This research was supported by the ARO Grant No. W911N-12-1-0263,  ISF Grant No. 1248/10 and the University of Rochester.

\end{document}